\documentclass[journal]{IEEEtran}

\ifCLASSINFOpdf
\else
   \usepackage[dvips]{graphicx}
\fi
\usepackage{url,amsmath,booktabs,multirow}
\usepackage{graphicx}
\usepackage{CJKutf8, color, tipa, amsmath,amssymb, balance,pifont}

\begin{document}

% \title{A Unified Recognition and Correction Model for Accented Speech with Large Language Models}
% \title{A Single Model for Recognizing and Correcting Accented Speech Using Large Language Models}
\title{MMGER: Multi-modal and Multi-granularity Generative Error Correction with LLM for Joint Accent and Speech Recognition}

% \author{First A. Author, \IEEEmembership{Fellow, IEEE}, Second B. Author, and Third C. Author, Jr., \IEEEmembership{Member, IEEE}
% \thanks{This paragraph of the first footnote will contain the date on which you submitted your paper for review. It will also contain support information, including sponsor and financial support acknowledgment. For example, ``This work was supported in part by the U.S. Department of Commerce under Grant BS123456.'' }
% \thanks{The next few paragraphs should contain the authors' current affiliations, including current address and e-mail. For example, F. A. $X_{Accent}$Author is with the National Institute of Standards and Technology, Boulder, CO 80305 USA (e-mail: author@boulder.nist.gov).}
% \thanks{S. B. Author, Jr., was with Rice University, Houston, TX 77005 USA. He is now with the Department of Physics, Colorado State University, Fort Collins, CO 80523 USA (e-mail: author@lamar.colostate.edu).}}

\author{Bingshen Mu, Yangze Li, Qijie Shao, Kun Wei, Xucheng Wan, Naijun Zheng, Huan Zhou, \\Lei Xie, \IEEEmembership{Senior Member, IEEE}
% \thanks{This paragraph of the first footnote will contain the date on which you submitted your paper for review. It will also contain support information, including sponsor and financial support acknowledgment. For example, ``This work was supported in part by the U.S. Department of Commerce under Grant BS123456.'' }
\thanks{Corresponding author: Lei Xie.}
\thanks{Bingshen Mu, Yangze Li, Qijie Shao, Kun Wei, and Lei Xie are with the ASLP Lab, School of Computer Science, Northwestern Polytechnical University, Xi’an 710129, China (e-mail: bsmu@mail.nwpu.edu.cn; yzli@mail.nwpu.edu.cn; qjshao@npu-aslp.org; ethanwei@mail.nwpu.edu.cn; lxie@nwpu.edu.cn).}
\thanks{Xucheng Wan, Naijun Zheng, and Huan Zhou are with the Huawei IIRC, Shenzhen 518129, China (e-mail: wanxucheng@huawei.com; zhengnaijun@huawei.com; zhou.huan@huawei.com).}}
% \author{Bingshen Mu, Xucheng Wan, Naijun Zheng, Huan Zhou, Lei Xie, \IEEEmembership{Senior Member, IEEE}
% % \thanks{This paragraph of the first footnote will contain the date on which you submitted your paper for review. It will also contain support information, including sponsor and financial support acknowledgment. For example, ``This work was supported in part by the U.S. Department of Commerce under Grant BS123456.'' }
% \thanks{Corresponding author: Lei Xie.}
% \thanks{Bingshen Mu and Lei Xie are with the ASLP Lab, School of Computer Science, Northwestern Polytechnical University, Xi’an 710129, China (e-mail: bsmu@mail.nwpu.edu.cn; lxie@nwpu.edu.cn).}
% \thanks{Xucheng Wan, Naijun Zheng, and Huan Zhou are with the Huawei IIRC, Shenzhen 518129, China (e-mail: wanxucheng@huawei.com; zhengnaijun@huawei.com; zhou.huan@huawei.com).}}

\markboth{Journal of \LaTeX\ Class Files, Vol. 14, No. 8, August 2015}
{Shell \MakeLowercase{\textit{et al.}}: Bare Demo of IEEEtran.cls for IEEE Journals}
\maketitle

\begin{abstract}
% Despite notable advancements in automatic speech recognition (ASR) technology, performance degradation when confronted with adverse conditions.
Despite notable advancements in automatic speech recognition (ASR), performance tends to degrade when faced with adverse conditions.
% Generative error correction (GER) leverages the outstanding text comprehension abilities of large language models (LLM), delivering impressive performance in ASR error correction, where N-best hypotheses provide valuable information for transcription prediction.
Generative error correction (GER) leverages the exceptional text comprehension capabilities of large language models (LLM), delivering impressive performance in ASR error correction, where N-best hypotheses provide valuable information for transcription prediction.
% However, GER suffers from issues such as fixed N-best hypotheses, inadequate utilization of acoustic information, and lack of specificity to the multi-accent scenario.
However, GER encounters challenges such as fixed N-best hypotheses, insufficient utilization of acoustic information, and limited specificity to multi-accent scenarios.
% However, GER encounters challenges such as fixed N-best hypotheses and insufficient utilization of acoustic information.
% The acoustic deviations between accent and standard pronunciation lead to linguistic discrepancies between hypotheses and transcriptions.
In this paper, we explore the application of GER in multi-accent scenarios.
% The acoustic differences between accented and standard pronunciations result in linguistic disparities between hypotheses and transcriptions.
% Although the multi-task learning framework for joint ASR and accent recognition (AR) has been proven effective in handling the multi-accent scenario, this framework only integrates acoustic and linguistic information through a shared encoder, limiting its performance when combined with GER.
% While the multi-task learning framework for simultaneous ASR and accent recognition (AR) has demonstrated efficacy in addressing the multi-accent scenario, it solely integrates acoustic and linguistic information via a shared encoder, constraining its effectiveness when combined with GER.
Accents represent deviations from standard pronunciation norms, and the multi-task learning framework for simultaneous ASR and accent recognition (AR) has effectively addressed the multi-accent scenarios, making it a prominent solution.
% In this work, we propose a unified recognition and correction model using multi-task ASR-AR learning, multimodal correction, and multi-granularity correction called AccentLLM.
In this work, we propose a unified ASR-AR GER model, named MMGER, leveraging multi-modal correction, and multi-granularity correction.
% Multi-task ASR-AR learning is utilized to provide dynamic 1-best hypotheses and accent embeddings.
Multi-task ASR-AR learning is employed to provide dynamic 1-best hypotheses and accent embeddings.
% Multimodal correction achieves fine-grained frame-level correction by force-aligning the acoustic features of speech with the corresponding character-level 1-best hypothesis sequence.
Multi-modal correction accomplishes fine-grained frame-level correction by force-aligning the acoustic features of speech with the corresponding character-level 1-best hypothesis sequence.
% Multi-granularity correction complements the global linguistic information by introducing regular 1-best hypotheses on top of fine-grained multimodal correction to achieve coarse-grained utterance-level correction.
Multi-granularity correction supplements the global linguistic information by incorporating regular 1-best hypotheses atop fine-grained multi-modal correction to achieve coarse-grained utterance-level correction.
% AccentLLM effectively addresses the shortcomings of GER and adapts LLM-based ASR error correction to the multi-accent scenario.
MMGER effectively mitigates the limitations of GER and tailors LLM-based ASR error correction for the multi-accent scenarios.
Experiments conducted on the multi-accent Mandarin KeSpeech dataset demonstrate the efficacy of MMGER, achieving a 26.72\% relative improvement in AR accuracy and a 27.55\% relative reduction in ASR character error rate, compared to a well-established standard baseline.
\end{abstract}
% \vspace{-9pt}

\IEEEpeerreviewmaketitle

\vspace{-9pt}
\section{Introduction}   
% \IEEEPARstart{A}{utomatic} Speech Recognition (ASR) has become increasingly crucial in modern society due to its ability to efficiently and accurately transcribe spoken language.
% As deep learning continues to advance, ASR has made significant progress, leading to substantial improvements in its performance. 
% However, ASR encounters numerous errors when dealing with speech variations caused by various factors such as background noise~\cite{mu2024automatic}, speaker accents~\cite{shao2023decoupling}, and speaker gender~\cite{boito2022study}.
\IEEEPARstart{A}{utomatic} Speech Recognition (ASR) has emerged as an increasingly crucial technology in modern society owing to its capability for the efficient and precise transcription of spoken language.
With the continuous advancement of deep learning, ASR has witnessed considerable progress in its efficacy.
Nevertheless, ASR encounters numerous errors when confronted with speech variations stemming from diverse factors such as background noise~\cite{mu2024automatic} and speaker accents~\cite{shao2023decoupling}.
% Various language model (LM) rescoring methods such as shallow fusion~\cite{kannan2018analysis}, deliberation~\cite{hu2020deliberation}, component fusion~\cite{shan2019component}, and cold fusion~\cite{sriram2017cold} have been widely applied in ASR decoding to enhance the linguistic acceptability of recognition results, thereby achieving a stable improvement in ASR performance.
% Various language model (LM) rescoring methods~\cite{kannan2018analysis, hu2020deliberation, shan2019component, sriram2017cold} have been widely applied in ASR decoding to enhance the linguistic acceptability of recognition results, thereby achieving a stable improvement in ASR performance.
Various language model (LM) rescoring methods~\cite{kannan2018analysis, hu2020deliberation, shan2019component, sriram2017cold} have been widely employed in ASR decoding to enhance the linguistic coherence of recognition results, thereby consistently enhancing ASR performance.
% Moreover, LM is also extensively employed in error correction tasks, utilizing the 1-best or N-best hypotheses generated by the ASR model~\cite{leng2021fastcorrect,mani2020asr, leng2023softcorrect, ma2023n}. 
Moreover, LM is also extensively utilized in ASR error correction tasks, leveraging either 1-best or N-best hypotheses generated by the ASR model~\cite{leng2021fastcorrect,mani2020asr, leng2023softcorrect, ma2023n}. 
% Leveraging the exceptional text comprehension capabilities of large language models (LLM), recent works propose a generative error correction (GER) framework to better mine the information in N-best hypotheses from ASR~\cite{chen2024hyporadise, chen2023generative}.
Leveraging the remarkable text comprehension capabilities of large language models (LLM), recent works have proposed a generative error correction (GER) framework to better mine the information in N-best hypotheses from ASR~\cite{chen2024hyporadise, chen2023generative}.
Through parameter-efficient LLM finetuning, GER demonstrates outstanding performance in learning the mapping from hypotheses to transcriptions, significantly outperforming typical LM rescoring methods. To facilitate GER in noise-robust scenarios, 
RobustGER~\cite{hu2024large} employs knowledge distillation (KD) to distill the real noise information from source speech into the language embedding GER to achieve language-space denoising, demonstrating remarkable noise-robustness.

% Although GER achieves excellent performance, it still exhibits some drawbacks.
While GER achieves remarkable performance, it still manifests certain limitations.
Firstly, GER is essentially a cascade framework, initially generating N-best hypotheses through speech foundation models such as WavLM~\cite{chen2022wavlm} and Whisper~\cite{radford2023robust}, then learning the mapping between these hypotheses and ground-truth transcriptions using LLM.
% The N-best hypotheses generated in this way are fixed and may limit the performance and generalization ability of the GER framework.
The N-best hypotheses generated in this way are fixed and potentially constrain the performance and generalization capability of the GER framework.
Secondly, GER relies solely on the linguistic information from the N-best hypotheses to predict transcriptions, lacking effective utilization of the acoustic information inherent in the speech.

%Moreover, it is noteworthy that the dataset HyPoradise~\cite{chen2024hyporadise} used in the GER consists of abundant pairs of ASR N-best hypotheses and transcription collected from diverse ASR corpus in prevalent speech domains.

% However, GER-based methods have not yet addressed the particularities of multi-accent scenarios.

In this paper, we explore the application of GER in challenging accented speech scenarios.
Accents represent deviations from standard pronunciation norms influenced by the speaker's educational background, geographical region, or native language~\cite{markl23_interspeech}.
% Recent studies suggest that multi-accent challenge can be addressed from both acoustic and linguistic aspects, with the multi-task learning framework for simultaneous ASR and accent recognition (AR) serving as a prevalent solution~\cite{jain18_interspeech, zhang21j_interspeech, toshniwal2018multilingual}.
Recent studies suggest that addressing the multi-accent challenge can involve both acoustic and linguistic perspectives, with the multi-task learning framework for joint ASR and accent recognition (AR) emerging as a prominent solution~\cite{jain18_interspeech, zhang21j_interspeech, toshniwal2018multilingual}.
This framework typically consists of a shared encoder and two branches, one for the ASR task and the other for the AR task.
On the one hand, as opposed to solely relying on acoustic information, simultaneously incorporating acoustic information and linguistic information from the ASR task can effectively enhance AR performance~\cite{zhang21j_interspeech,shao22b_interspeech, shi2021accented}.
% On the other hand, incorporating accent information from the AR task into the ASR task enables ASR to adapt to pronunciation or linguistic variations, thereby enhancing ASR performance on accented speech~\cite{jain18_interspeech,qian2022layer, imaizumi2020dialect}.
On the other hand, incorporating accent information from the AR task into ASR enables adaptation to pronunciation or linguistic variations, thereby enhancing ASR performance with accented speech~\cite{jain18_interspeech,qian2022layer, imaizumi2020dialect}.

% Inspired by RobustGER~\cite{hu2024large}, an intuitive approach to tailor GER for the multi-accent scenario is to integrate the multi-task ASR-AR framework with LLM, allowing LLM to map the dynamic 1-best hypotheses and accent embeddings predicted by the multi-task ASR-AR framework to transcriptions.
% However, this approach only merges acoustic and linguistic information through a shared encoder.
% However, this approach only combines acoustic and linguistic information via a shared encoder.
% Furthermore, LLM only utilizes linguistic information from the 1-best hypotheses to correct errors, lacking the utilization of acoustic information.
% Additionally, LLM solely employs linguistic information from the 1-best hypotheses to correct errors, neglecting the utilization of acoustic information.
% Our examination in Table~\ref{tab:table1} also demonstrates these limitations. 

To this end, we propose a unified ASR-AR GER model leveraging multi-modal correction and multi-granularity correction, named \textit{MMGER}.
Multi-task ASR-AR learning is employed to generate dynamic 1-best hypotheses and accent embeddings, alleviating the constraint of using fixed N-best hypotheses in GER and facilitating the model in adapting to the acoustic and linguistic variations of particular accents.
% The deviation of accents from standard pronunciation in the acoustic aspect leads to ASR hypotheses of accent speech deviating from the transcription in the linguistic aspect.
The acoustic differences between accented and standard pronunciations result in linguistic disparities between hypotheses and transcriptions.
After force-aligning the acoustic features of accented speech with its character-level 1-best hypothesis sequence, we derive a multi-modal representation of the accent.
\textit{Multi-modal} correction achieves fine-grained frame-level correction by leveraging the multi-modal representation of accents.
Since frame-level multi-modal representations diminish the global linguistic information in speech, we introduce \textit{multi-granularity} correction atop multi-modal correction, achieving coarse-grained utterance-level correction by leveraging the global linguistic information contained in the regular 1-best hypotheses.
% As a result, LLM can effectively improve error correction performance on accented speech by using multi-task ASR-AR learning, multimodal correction, and multi-granularity correction.
Experiments conducted on the multi-accent Mandarin KeSpeech~\cite{tang2021kespeech} dataset demonstrate the effectiveness of MMGER, achieving a 26.72\% AR accuracy (ACC) relative improvement and a 27.55\% ASR character error rate (CER) relative reduction over a well-established baseline.
% It is noteworthy that AccentLLM achieves state-of-the-art results in terms of CER on the KeSpeech dataset.
Impressively, MMGER attains state-of-the-art CER on the multi-accent Mandarin KeSpeech dataset.

\vspace{-9pt}
\section{Proposed Approach}
\vspace{-9pt}
\subsection{Overview}
\begin{figure}[ht]
  \centering
  \includegraphics[width=\linewidth]{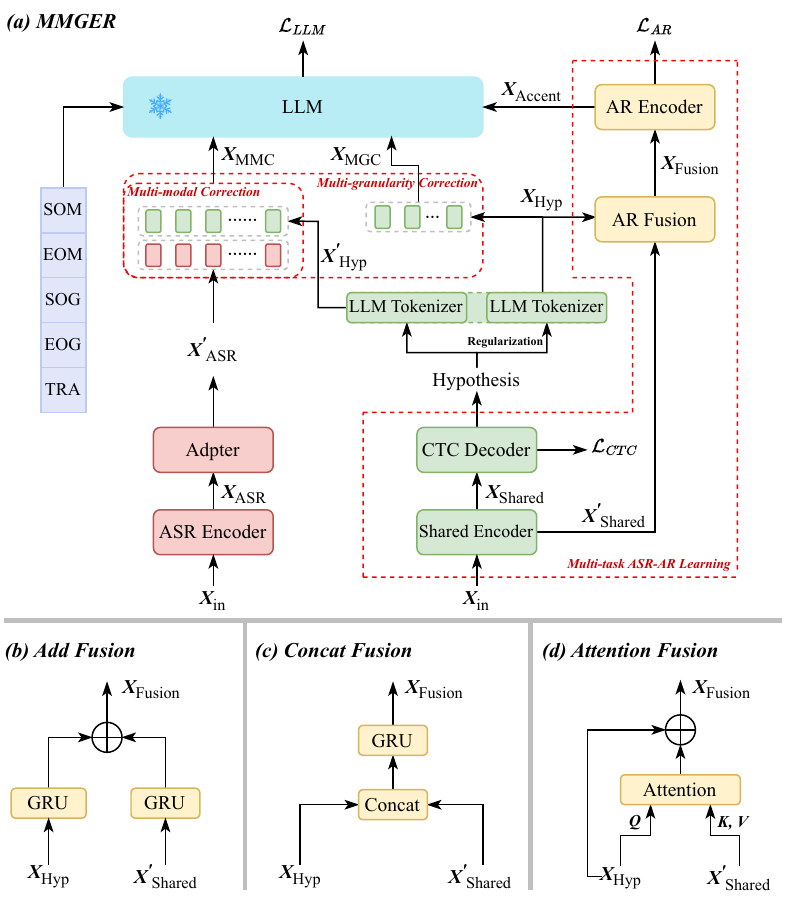}
  \caption{(a) An overview of our proposed MMGER. (b) The architecture of the additive fusion (Add Fusion). (c) The architecture of the concatenation fusion (Concat Fusion). (d) The architecture of the Attention Fusion.}
  \vspace{-9pt}
  \label{fig:MMGER}
\end{figure}
% \textless$\left | transcribe \right |$\textgreater 
As depicted in Fig.~\ref{fig:MMGER}(a), MMGER comprises a multi-task ASR-AR learning module, a multi-modal correction module, a multi-granularity correction module, and a frozen LLM. Motivated by Whisper~\cite{radford2023robust}, we utilize five special tokens, namely \textless\textbar SOM\textbar\textgreater, \textless\textbar EOM\textbar\textgreater, \textless\textbar SOG\textbar\textgreater, \textless\textbar EOG\textbar\textgreater, and \textless\textbar TRA\textbar\textgreater, to represent the start and end of the multi-modal correction, the start and end of the multi-granularity correction, and transcription respectively.
Additionally, we explore three fusion schemes for incorporating acoustic and linguistic information for the AR task within the multi-task ASR-AR learning module: additive fusion, concatenation fusion, and attention fusion.
We will elaborate on the details of all components of MMGER in the subsequent subsections.
\vspace{-9pt}
\subsection{Multi-task ASR-AR Learning}
The multi-task ASR-AR learning module comprises a shared encoder, a connectionist temporal classification (CTC)~\cite{graves2006connectionist} decoder, an AR fusion, and an AR encoder. 
The objective of the ASR task is to provide dynamically changing 1-best hypotheses. 
Specifically, the shared encoder extracts acoustic information from the speech features $\mathbf{X}_{\mathrm{in}}$, and the CTC decoder generates the 1-best hypotheses of the speech through CTC greedy search. 
As the parameters of the shared encoder and CTC decoder are updated during training, the 1-best hypotheses undergo dynamic changes.
% The computation of the shared encoder output $\mathbf{X}_{\mathrm{Shared}}$ and the loss of the ASR task can be denoted as
The loss of the ASR task can be denoted as
\begin{equation}
    \left\{
        \begin{aligned}
            &\mathbf{X}_{\mathrm{Shared}} = \text{Encoder}_{\text{Shared}}(\mathbf{X}_{\mathrm{in}}),\\
            &\mathcal{L}_{\mathrm{CTC}} = -\log P_{\mathrm{CTC}}(\mathbf{Y}_{\mathrm {CTC}} \mid \text{Decoder}_{\text{CTC}}(\mathbf{X}_{\mathrm{Shared}}))\\
        \end{aligned}
    \right.
\end{equation}
where $\mathbf{X}_{\mathrm{in}}$ is the FBank feature, $\mathbf{Y}_{\mathrm {CTC}}$ is the labels tokenized by the CTC tokenizer, and the CTC decoder is a linear projector used to map the feature dimension of $\mathbf{X}_{\mathrm{Shared}}$ to the size of the CTC vocabulary.

Additionally, the objective of the AR task is to deliver utterance-level accent embeddings for the speech. 
% The higher the accuracy of the AR task, the better the quality of the accent embeddings.
From our perspective, accent embeddings should be derived jointly from both acoustic and linguistic information. 
% Therefore, we regard $\mathbf{X}_{\mathrm{Shared}}^{'}$ as the acoustic information and $\mathbf{X}_{\mathrm{Hyp}}$ as the linguistic information, exploring three different fusion methods to improve AR performance.
With this aim in mind, multiple intermediate features from the shared encoder are accumulated as acoustic information $\mathbf{X}_{\mathrm{Shared}}^{'}$, while the tokenized hypothesis from LLM serves as linguistic information $\mathbf{X}_{\mathrm{Hyp}}$.
The $\mathbf{X}_{\mathrm{Shared}}^{'}$ and $\mathbf{X}_{\mathrm{Hyp}}$ can be calculated as:
\begin{equation}
    \left\{
        \begin{aligned}
            &\mathbf{X}_{\mathrm{Shared}}^{'} = \text{Concat}(\text{Encoder}_{\text{Shared}}(\mathbf{X}_{\mathrm{in}})^{i,j,k,l}),\\
            &\mathbf{X}_{\mathrm{Hyp}} = \text{Tokenizer}_{\text{LLM}}(\text{Regularization}(\text{Hypothesis})),\\
        \end{aligned}
    \right.
\end{equation}
where superscript ${i,j,k,l}$ denote intermediate output positions within the shared encoder and \textit{Regularization} refers to remove blanks and duplicated tokens.
% Three schemes are investigated to merge acoustic and linguistic information: additive, concatenation, and attention fusion.
Fig.~\ref{fig:MMGER}(b) illustrates the details of the additive fusion (Add Fusion), where $\mathbf{X}_{\mathrm{Hyp}}$ and $\mathbf{X}_{\mathrm{Shared}}^{'}$ are independently processed through a Gated Recurrent Unit (GRU)~\cite{cho2014learning} and then added together to obtain $\mathbf{X}_{\mathrm{Fusion}}$. The output of the Add Fusion is calculated as
\begin{equation}
    \mathbf{X}_{\mathrm{Fusion}} = \text{Add}(\text{GRU}_{\text{Hyp}}(\mathbf{X}_{\mathrm{Hyp}}),\text{GRU}_{\text{Shared}}(\mathbf{X}_{\mathrm{Shared}}^{'}))
\end{equation}
where the final hidden states of two GRUs are utterance-level embeddings that eliminate the length dimension of $\mathbf{X}_{\mathrm{Shared}}^{'}$ and $\mathbf{X}_{\mathrm{Hyp}}$.
Fig.~\ref{fig:MMGER}(c) illustrates the details of the concatenation fusion (Concat Fusion), where $\mathbf{X}_{\mathrm{Shared}}^{'}$ and $\mathbf{X}_{\mathrm{Hyp}}$ are initially concatenated along the length dimension and then processed through a GRU to yield $\mathbf{X}_{\mathrm{Fusion}}$.
The output of the Concat Fusion is calculated as
\begin{equation}
    \mathbf{X}_{\mathrm{Fusion}} = \text{GRU}(\text{Concat}(\mathbf{X}_{\mathrm{Hyp}},\mathbf{X}_{\mathrm{Shared}}^{'}))
\end{equation}
Fig.~\ref{fig:MMGER}(d) illustrates the details of the Attention Fusion, $\mathbf{X}_{\mathrm{Hyp}}$ acts as the query, $\mathbf{X}_{\mathrm{Shared}}^{'}$ as both the key and value, and the outcome of attention calculation is then residual connected with $\mathbf{X}_{\mathrm{Hyp}}$ to obtain $\mathbf{X}_{\mathrm{Fusion}}$.
The output of the Attention Fusion is calculated as
\begin{equation}
    \resizebox{0.9\hsize}{!}{$
    \mathbf{X}_{\mathrm{Fusion}} = \text{Add}\left (  \text{Softmax} \left (\frac{\mathbf{X}_{\mathrm{Hyp}}\left (\mathbf{X}_{\mathrm{Shared}}^{'}\right )^{T}}{\sqrt{D}}  \right )\mathbf{X}_{\mathrm{Shared}}^{'}, \mathbf{X}_{\mathrm{Hyp}}\right )
    $}
\end{equation}
where $D$ is the feature dimension of $\mathbf{X}_{\mathrm{Shared}}^{'}$.
The loss of the AR task is calculated as
\begin{equation}
    \resizebox{0.9\hsize}{!}{$
    \mathcal{L}_{\mathrm{AR}} = \text{CrossEntropy}(P(\mathbf{Y}_{\mathrm{AR}}\mid\text{Encoder}_{\text{AR}}(\mathbf{X}_{\mathrm{Fusion}})), \mathbf{Y}_{\mathrm{AR}})$}
\end{equation}
where $\mathbf{Y}_{\mathrm{AR}}$ is the accent label.
$\mathbf{X}_{\mathrm{Accent}}$ is an intermediate result of the AR encoder.
The higher the accuracy achieved in the AR task, the better the quality of the accent embeddings.
\vspace{-9pt}
\subsection{Multi-modal Correction}
% The deviation of accents from standard pronunciation in the acoustic aspect leads to ASR hypotheses of accent speech deviating from the transcription in the linguistic aspect.
The acoustic differences between accented and standard pronunciations result in linguistic disparities between hypotheses and transcriptions.
The acoustic deviation and linguistic deviation together constitute the multi-modal representation of accents.
% To prevent interference with the multi-task ASR-AR module, we utilize an ASR encoder to extract the speech embedding as the acoustic deviation, while avoiding using the speech embedding extracted by the shared encoder.
To mitigate interference with the multi-task ASR-AR module, we utilize another ASR encoder to extract the speech embedding as the acoustic deviation, while refraining from using the speech embedding extracted by the shared encoder.
Inspired by Qwen-audio~\cite{chu2023qwen}, we employ an adapter to align the speech embedding extracted by the ASR encoder to the text modality required by LLM.
As a result, the speech embedding $\mathbf{X}_{\mathrm{ASR}}^{'}$ is the acoustic deviation.
Furthermore, we employ the character-level 1-best hypothesis sequence to force-align linguistic deviation with acoustic deviation without removing blank and duplicate tokens. 
We individually process each token in this hypothesis sequence through the LLM tokenizer to obtain embeddings for each token. 
Subsequently, we concatenate these embeddings together to create the linguistic deviation $\mathbf{X}_{\mathrm{Hyp}}^{'}$, which matches the length of the acoustic deviation.
Concatenating the acoustic deviation and linguistic deviation along the feature dimension yields the frame-level multi-modal representation of the accent, enabling fine-grained multi-modal correction.
The multi-modal representation $\mathbf{X}_{\mathrm{MMC}}$ is denoted as
\begin{equation}
    \left\{
        \begin{aligned}
            &\mathbf{X}_{\mathrm{ASR}}^{'} = \text{Adpter}(\text{Encoder}_{\text{ASR}}(\mathbf{X}_{\mathrm{in}})) ,\\
            &\mathbf{X}_{\mathrm{Hyp}}^{'} = \text{Tokenizer}_{\text{LLM}}(\text{Hypothesis}),\\
            &\mathbf{X}_{\mathrm{MMC}} = \text{Concat}(\mathbf{X}_{\mathrm{ASR}}^{'}, \mathbf{X}_{\mathrm{Hyp}}^{'}),\\
        \end{aligned}
    \right.
\end{equation}
where $\mathbf{X}_{\mathrm{ASR}}^{'}$ and $\mathbf{X}_{\mathrm{Hyp}}^{'}$ are with the same length.
\vspace{-9pt}
\subsection{Multi-granularity Correction}
Since the linguistic deviation in the multi-modal representation of the accent is derived from each token in the 1-best hypothesis sequence through the LLM tokenizer, this leads to the loss of global linguistic information in the linguistic deviation.
However, global linguistic information is crucial in ASR error correction.
Therefore, we build upon fine-grained frame-level multi-modal correction by leveraging the global utterance-level linguistic information contained in $\mathbf{X}_{\mathrm{MGC}}$ which is generated by 1-best hypothesis after regularization through the LLM tokenizer to accomplish coarse-grained utterance-level error correction.
The combination of $\mathbf{X}_{\mathrm{MMC}}$ and $\mathbf{X}_{\mathrm{MGC}}$ is the multi-granularity correction.
The $\mathbf{X}_{\mathrm{MGC}}$ can be calculated as
\begin{equation}
    \mathbf{X}_{\mathrm{MGC}} = \text{Tokenizer}_{\text{LLM}}(\text{Regularization}(\text{Hypothesis}))
\end{equation}
where $\mathbf{X}_{\mathrm{Hyp}}$ and $\mathbf{X}_{\mathrm{MGC}}$ are identical.
\vspace{-9pt}
\subsection{Frozen LLM}
The prompt format for the frozen LLM adheres to the following rule:
``\textless\textbar SOM\textbar\textgreater\,$\mathbf{X}_{\mathrm{MMC}}$\,\textless\textbar EOM\textbar\textgreater\,\textless\textbar SOG\textbar\textgreater\,$\mathbf{X}_{\mathrm{MGC}}$ \,\textless\textbar EOG\textbar\textgreater\,\textless\textbar TRA\textbar\textgreater\,$\mathbf{X}_{\mathrm{Accent}}$ \,\textless Labels\textgreater'', where the five special tokens are trainable embeddings and \textless Labels\textgreater\, represents the corresponding transcription.
Through end-to-end joint training, leveraging the excellent text comprehension capabilities of LLM, we accomplish accent-specific multi-modal correction and multi-granularity correction by way of taking the frame-level accent multi-modal representation, regular 1-best hypotheses containing global linguistic information and accent embeddings as inputs to the frozen LLM.
% Through this approach, LLM can learn specific linguistic information related to accents in the hypotheses and subsequently refine these hypotheses based on both acoustic information and linguistic information.

Finally, the total loss of the MMGER consists of the LLM loss $\mathcal{L}_{\mathrm{LLM}}$, the CTC loss $\mathcal{L}_{\mathrm{CTC}}$, and the AR loss $\mathcal{L}_{\mathrm{AR}}$, which can be formulated as
\begin{equation}
    \mathcal{L}_{\mathrm{MMGER}} = \mathcal{L}_{\mathrm{LLM}} + \lambda\mathcal{L}_{\mathrm{CTC}} + \mu\mathcal{L}_{\mathrm{AR}}
\end{equation}
where $\lambda$ and $\mu$ are tunable hyperparameters.
\vspace{-9pt}
\section{Experiments}
\begin{table*}[th]
    \caption{The ACC (\%) and CER (\%) Results of Our Proposed Approaches on The KeSpeech Dev and Test Datasets. Reported ASR CER Results is In The Following Format: CER on All The Speech / CER on The Accented Speech.}
    \label{tab:table1}
    \centering
\begin{tabular}{lccccccc}
\toprule
\multirow{2}{*}{\textbf{ID}} & \multirow{2}{*}{\textbf{Multi-modal Correction}} & \multirow{2}{*}{\textbf{Multi-granularity Correction}} & \multirow{2}{*}{\textbf{AR Fusion}} & \multicolumn{2}{c}{\textbf{AR ACC (\%)}} & \multicolumn{2}{c}{\textbf{ASR CER(\%)}} \\ \cmidrule{5-8} 
                    &                                        &                                               &                            & \textbf{Dev}            & \textbf{Test}           & \textbf{Dev}            & \textbf{Test}           \\ \midrule
A1                  & \ding{52}                                      & \ding{52}                                             & Acoustic Only              & 88.21          & 77.39          & 4.64 / 8.88           & 7.80 / 9.21           \\
A2                  & \ding{52}                                      & \ding{52}                                             & Linguistic Only            & 88.15          & 26.24          & 5.01 / 9.10           & 8.20 / 9.66           \\
A3                  & \ding{52}                                      & \ding{52}                                             & Add Fusion                 & 90.12          & 77.47          & \textbf{4.50 / 8.35}           & \textbf{7.52 / 8.90}           \\
A4                  & \ding{52}                                      & \ding{52}                                             & Concat Fusion              & 90.17          & 77.54          & 4.60 / 8.77           & 7.62 / 8.99           \\
A5                  & \ding{52}                                      & \ding{52}                                             & Attention Fusion           & \textbf{90.33}          & \textbf{77.61}          & 4.57 / 8.53           & 7.66 / 9.04           \\ \midrule
G1                  & \ding{56}                                      & Coarse-grained Only                           & Add Fusion                 & 89.65          & 76.91          & 5.95 / 10.32           & 9.44 / 11.03          \\
G2                  & \ding{52}                                      & Fine-grained Only                             & Add Fusion                 & 89.40          & 76.37          & 5.13 / 10.02          & 8.56 / 10.16           \\ \midrule
M1                  & Acoustic Only                          & \ding{52}                                             & Add Fusion                 & 89.16          & 76.89          & 4.80 / 9.14           & 7.82 / 9.20           \\
M2                  & Linguistic Only                        & \ding{52}                                             & Add Fusion                 & 90.06          & 76.94          & 6.12 / 12.03           & 9.21 / 10.74           \\ \bottomrule
\vspace{-9pt}
\end{tabular}
\end{table*}
\subsection{Experimental Setup}
We conduct experiments using the KeSpeech~\cite{tang2021kespeech} dataset to evaluate the effectiveness of our proposed MMGER.
KeSpeech involves 1,542 hours of speech recorded by 27,237 speakers in 34 cities in China, and the pronunciation includes standard Mandarin and eight major accented Mandarin.

MMGER is implemented with the WeNet~\cite{yao2021wenet} toolkit. 
The ASR encoder is a 12-layer Conformer~\cite{gulati20_interspeech} where feed-forward networks (FFN) have 2048 dimensions, multi-head self-attentions (MHSA) have 512 dimensions, and the attention head is 8.  
The adapter is a 2-layer Transformer~\cite{vaswani2017attention} where the attention head is 4 and the dimensions of FFN and MHSA are respectively set to 2048 and 1024. 
The shared encoder is a 12-layer Conformer where the attention head is 4 and the dimensions of FFN and MHSA are respectively set to 2048 and 256. 
The AR encoder is a 3-layer linear projector with the GELU~\cite{hendrycks2016gaussian} activation function.
% The ASR encoder and CTC encoder are initialized with two well-trained ASR models that are trained on the Wenetspeech~\cite{zhang2022wenetspeech} dataset and the KeSpeech dataset, respectively.
The CTC decoder is a linear projector.
We use Qwen-7B~\cite{bai2023qwen} as our frozen LLM and the dimensions of the five special tokens and accent embeddings are 4096, and the hyperparameters of $\lambda$ and $\mu$ are 0.3.

\vspace{-9pt}
\subsection{Results and Discussion}
% \usepackage{multirow}
% \usepackage{multirow}

Table~\ref{tab:table1} presents the AR ACC and ASR CER results of MMGER on the KeSpeech Dev and Test datasets.
According to~\cite{tang2021kespeech}, the AR ACC averages six accents, excluding Northeastern and Beijing Mandarin.
Experiments A1 to A5 demonstrate the impact of different fusion methods for the AR task within the multi-task ASR-AR learning module on the performance of MMGER.
The poor quality of accent embeddings generated solely from unimodal information leads to a degradation in the performance of MMGER.
While the better quality of accent embeddings generated from multi-modal information helps MMGER achieve better performance, different fusion methods do not significantly impact the final performance of MMGER.
Furthermore, experiments G1 and G2 investigate the performance of multi-granularity correction, indicating that performing only single-granularity correction would adversely affect the performance of MMGER.
Finally, experiments M1 and M2 explore the significance of multi-modal correction, highlighting the crucial role of multi-modal accent representations in error correction.

Table~\ref{tab:table2} presents the AR ACC and ASR CER results of MMGER and other models on the KeSpeech Dev and Test datasets.
The original GER method outperforms the ASR model consisting of the conformer encoder and the hybrid CTC/Attention decoder with transformer-based LM rescoring (ASR Baseline + LM) but falls short compared to DIMNet with LM rescoring (DIMNet + LM).
Inspired by Qwen-audio, we attempt to replace the decoders of traditional ASR and multi-task ASR-AR learning network with Qwen-7B LLM (Conformer-LLM and MJTR-LLM) and achieve performance similar to GER.
MMGER achieves the best performance compared to the original GER and LLM-based ASR.
\begin{table}[]
    \caption{The ACC (\%) and CER (\%) Results of Our Proposed Model Compared With Other Models.}
    \label{tab:table2}
    \centering
\begin{tabular}{lcccc}
\toprule
\multirow{2}{*}{\textbf{Model}} & \multicolumn{2}{c}{\textbf{AR ACC (\%)}} & \multicolumn{2}{c}{\textbf{ASR CER (\%)}} \\ \cmidrule{2-5} 
                       & \textbf{Dev}        & \textbf{Test}       & \textbf{Dev}        & \textbf{Test}       \\ \midrule
KeSpeech Baseline~\cite{tang2021kespeech}      & -          & 61.13      & -          & 10.38      \\
ASR Baseline + LM      & -          & -          & 5.95       & 9.39       \\
Conformer-LLM          & -          & -          & 6.10       & 9.02       \\
MJTR~\cite{zhang21j_interspeech}                   & 65.25      & 67.00      & 7.79       & 12.77      \\
MJTR-LLM               & 64.10      & 65.81      & 6.10       & 9.99       \\
GER~\cite{chen2024hyporadise}                    & -          & -          & 5.98       & 9.35       \\
DIMNet + LM~\cite{shao2023decoupling}            & 80.06      & \textbf{78.57}      & 5.71       & 8.87       \\
MMGER              & \textbf{90.12}      & 77.47      & \textbf{4.50}       & \textbf{7.52}       \\ \bottomrule
\vspace{-9pt}
\end{tabular}
\end{table}
% \vspace{-9pt}
\begin{CJK}{UTF8}{gbsn}
    \begin{table}[h]
        \caption{Case Study of MMGER. These Cases are Selected from The KeSpeech Test Dataset, Including Lan-Yin Mandarin and Northeastern Mandarin.}
        \label{tab:table3}
        \centering
    \begin{tabular}{lll}
    \toprule
    \textbf{Accent}                                            & \textbf{Method}            & \textbf{Utterance}                       \\ \midrule
    \multirow{6}{*}{Lan-Yin}                          & CTC Greedy Search & \textcolor{red}{外/ua\textipa{I}4/} 间 \textcolor{red}{买/ma\textipa{I}3/} 盘 行 情      \\
                                                      & MMGER         & \textcolor{green}{晚/wan3/} 间 \textcolor{green}{美/me\textipa{I}3/} 盘 行 情      \\
                                                      & CTC Greedy Search & \textcolor{red}{更/k\textipa{GN}1/} 茎 \textcolor{red}{丽/li4/} 的 蔬 菜 \\
                                                      & MMGER         & \textcolor{green}{根/k\textipa{@}n1/} 茎 \textcolor{green}{类/le\textipa{I}4/} 的 蔬 菜 \\
                                                      & CTC Greedy Search & 其 中 酒 \textcolor{red}{税/\textipa{\:s}we\textipa{I}4/} 的 费 用        \\
                                                      & MMGER         & 其 中 酒 \textcolor{green}{水/\textipa{\:s}we\textipa{I}3/} 的 费 用        \\ \midrule
    \multicolumn{1}{c}{\multirow{6}{*}{Northeastern}} & CTC Greedy Search & 船 舶 过 \textcolor{red}{杂/tsa2/} 秩 序     \\
    \multicolumn{1}{c}{}                              & MMGER         & 船 舶 过 \textcolor{green}{闸/\textipa{\:t\:s}a2/} 秩 序     \\
    \multicolumn{1}{c}{}                              & CTC Greedy Search & 自 主 \textcolor{red}{从/ts$^{h}$\textipa{UN}2/} 高 端   \\
    \multicolumn{1}{c}{}                              & MMGER         & 自 主 \textcolor{green}{冲/\textipa{\:t\:s}$^{h}$\textipa{UN}1/} 高 端    \\
    \multicolumn{1}{c}{}                              & CTC Greedy Search & 受 到 \textcolor{red}{水/\textipa{\:s}we\textipa{I}3/} 害 的 个 人 \\
    \multicolumn{1}{c}{}                              & MMGER         & 受 到 \textcolor{green}{损/su\textipa{@}n3/} 害 的 个 人 \\ \bottomrule
    \vspace{-18pt}
    \end{tabular}
    \end{table}
\end{CJK}
\vspace{-9pt}
\subsection{Case Study}
Table~\ref{tab:table3} presents a case study of Lan-Yin Mandarin and Northeastern Mandarin to demonstrate the effectiveness of MMGER.
The red font indicates incorrect characters generated by CTC greedy search, whereas the green font signifies the corrected characters by MMGER.
To visually illustrate the influence of accents, we annotate characters with International Phonetic Alphabet (IPA) symbols.
The distinctions between Lan-Yin Mandarin and standard Mandarin mainly pertain to tone, nasal sounds, and final pronunciation, while the disparities between Northeastern Mandarin and standard Mandarin primarily involve initial pronunciation.
This case study illustrates that MMGER can perform multi-modal and multi-granularity correction for specific accents.
\vspace{-9pt}
\section{Conclusion}
In this work, we propose an ASR-AR solution for multi-accent scenarios named MMGER, which serves as a unified multi-task ASR-AR learning GER model using multi-modal correction and multi-granularity correction.
Experiments on the multi-accent Mandarin KeSpeech dataset demonstrate the effectiveness of MMGER.
\clearpage
\bibliographystyle{IEEEbib}
\balance
\bibliography{refs}

\end{document}